# 3D periodic dielectric composite homogenization based on the Generalized Source Method

[1]A.A. Shcherbakov, [2]A.V. Tishchenko

[1]Moscow Institute of Physics and Technology, Dolgoprudny, Russia

[2]University of Lyon, Saint-Etienne, France

## Abstract

The article encloses a new Fourier space method for rigorous optical simulation of 3D periodic dielectric structures. The method relies upon rigorous solution of Maxwell's equations in complex composite structures by the Generalized Source Method. Extremely fast GPU enabled calculations provide a possibility for an efficient search of eigenmodes in 3D periodic complex structures on the basis of rigorously obtained resonant electromagnetic response. The method is applied to the homogenization problem demonstrating anisotropic dielectric tensor retrieval.

## Introduction

Artificial dielectric 3D periodic structures with sub-wavelength periods are promising candidates for replacing natural materials in various optical devices [1-4]. These structures operate in a single mode regime and provide benefits of engineering effective permittivity tensor for specific applications. Currently, 3D periodic photonic crystals with complex period filling are most efficiently produced by the Direct Laser Writing technique [5]. Despite the reduction of periods towards hundreds of nanometers requires an additional experimental and technological effort, current success in fabrication needs a support in form of reliable effective dielectric tensor simulation and optimization methods. Since the most prominent modification of the effective refractive indices is achieved for high-contrast structures of periods comparable with operation wavelengths, zero or small wavevector length limit approximations (e.g., [6-9]) provide insufficient accuracy [10] or lead to divergent series, and a rigorous solution of Maxwell's equations is required at some step of a calculation algorithm.

Rigorous numerical solutions to the homogenization problem for optical 3D periodic structures can be obtained by various methods [11,12]. In case of infinite bulk dielectric structures the Fourier decomposition techniques of the Maxwell's equations are widely used for photonic crystal band diagram calculation and appear to be among the most efficient approaches [13-15]. This is analogous to the grating diffraction theory where the Fourier Modal Method (FMM) and its numerous variations [16,17] is one of the most widespread tools for diffraction analysis. A powerful alternative to the FMM, the Generalized Source Method (GSM), was recently proposed in [18,19]. The GSM for rigorous grating diffraction calculation was demonstrated to outperform the FMM exhibiting a linear numerical complexity relative to the Fourier orders number. A specific GSM feature is that the method avoids solving the matrix eigenvalue problem while replacing the latter with solution of a linear equation system. Due to the Toeplitz structure of involved matrices such system can be efficiently solved by the Krylov subspace iterative techniques.

In this work we formulate the GSM for 3D periodic structures, and focus on the homogenization problem in complex anisotropic dielectric structures. The paper is organized as follows. First, we define a homogenization problem to solve. Second, we describe application of the GSM to rigorous diffraction calculation in 3D periodic dielectric composites. Albeit the resulting equations are the same to those obtained from conventional treatment of the Maxwell's equations in the differential form, here we keep sources with necessity and formulate our method as primarily integral one starting from a solution written via dyadic Green's function. This description is followed by representation of our approach to effective dielectric tensor components retrieval, which replaces conventional search of matrix eigenvalues [14]. The article ends up with numerical examples and discussion of a perspective work.

## Artificial dielectric structures

Consider simulation of linear optical response in 3D periodic dielectric structures. An example of such structure is shown in Fig. (1), being a 3D periodic lattice of spheres. Denote structure periods as $\Lambda_\alpha$, $\alpha = 1,2,3$. Despite the below formulation is rigorous without any assumptions on $\Lambda_\alpha$ values, in numerical examples we will suppose these periods to be smaller than wavelength $\lambda$ to support the 0-th propagating diffraction order only.

Since we aim at performing our analysis within the optical band, all materials are considered to be nonmagnetic with vacuum permeability $\mu_0$. Optical permittivity of dielectrics usually varies in range from 1 to ~10. Dielectric permittivity of a composite medium is supposed to be infinitely periodic $\varepsilon(\mathbf{r}) = \varepsilon\left(\mathbf{r} + \sum_{\alpha=x,y,z} m_\alpha \mathbf{a}_\alpha\right)$ with $|\mathbf{a}_\alpha| = \Lambda_\alpha$ and integers $m_\alpha$. Rigorous simulation of such 3D periodic structure requires numerical solution of the time-harmonic Maxwell's equations

$$\nabla \times \mathbf{E} = i\omega\mu_0 \mathbf{H}$$
$$\nabla \times \mathbf{H} = \mathbf{J} - i\omega\varepsilon \mathbf{E} \qquad (1)$$

for given permittivity distribution $\varepsilon(\mathbf{r}, \omega)$ supposing temporal variations of all fields and sources to be described by factor $\exp(-i\omega t)$. Equations (1) should be solved under boundary conditions consisting in continuity of the tangential electric field and the normal electric displacement components relative to interfaces between different media.

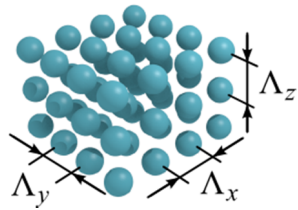

Figure 1. Example of a 3D periodic dielectric structure. Each period of the structure is supposed to be small enough comparable with light wavelength $\lambda$: $\Lambda_\alpha < \lambda$, $\alpha = 1,2,3$, to support at most two propagating modes in each propagation direction, i.e. to behave as an effective medium.

## Generalized source method for 3D periodic structures

To apply the GSM to 3D diffraction simulation and to resolve Eqs. (1), we briefly review the method rationale [20]. The GSM sequentially searches for solutions of Eqs. (1) in the following two steps. First, one chooses a basis problem to replace $\varepsilon(\mathbf{r})$ with some basis function $\varepsilon_b(\mathbf{r})$. The latter should support a known solution to Eqs. (1) for any given source distribution $\mathbf{J}$: $\mathbf{E} = \aleph_b(\mathbf{J})$. At the second step, one goes back to the initial diffraction problem by introducing a generalized source $\mathbf{J}^{gen} = -i\omega[\varepsilon(\mathbf{r}) - \varepsilon_b(\mathbf{r})]\mathbf{E}$ and substituting it into the basis solution in order to get a self-consistent implicit equation

$$\mathbf{E} = \mathbf{E}_{inc} + \aleph_b[-i\omega(\varepsilon - \varepsilon_b)\mathbf{E}], \qquad (2)$$

where $\mathbf{E}_{inc}$ is an initial field excited by real sources in a basis medium with $\varepsilon_b(\mathbf{r})$. A proper discretization scheme allows for reducing Eq. (2) to a set of linear algebraic equations which can be solved by efficient iterative methods.

Equation system (1) written in a basis medium with constant permittivity $\varepsilon_b = const$ gives rise to the well-known implicit integral equation [21]

$$E_\alpha(\mathbf{r}) = E_\alpha^{inc}(\mathbf{r}) + i\omega\mu_0 \sum_{\beta=1,2,3} \int_{V'} \left(\delta_{\alpha\beta} + \frac{1}{k_b^2}\frac{\partial}{\partial x_\alpha}\frac{\partial}{\partial x_\beta}\right) \frac{\exp(ik_b|\mathbf{r}-\mathbf{r}'|)}{4\pi|\mathbf{r}-\mathbf{r}'|} J_\beta(\mathbf{r}')d^3r', \qquad (3)$$

where $k_b^2 = \omega^2\varepsilon_b\mu_0$, $\delta_{\alpha\beta}$ is the Kronecker symbol, and the multiplier before source $J_\beta(\mathbf{r}')$ is the free space dyadic Green function [21]. Substitution of the generalized current into Eq. (3) yields a self-consistent linear equation which in the discreet 3D Fourier basis:

$$E_{\alpha\mathbf{m}} = E_{\alpha\mathbf{m}}^{inc} + \sum_{\beta=1,2,3} \frac{k_b^2\delta_{\alpha\beta} - k_{\alpha\mathbf{m}}k_{\beta\mathbf{m}}}{k_\mathbf{m}^2 - k_b^2} \sum_{\mathbf{m}'} \left(\frac{\Delta\varepsilon}{\varepsilon_b}\right)_{\mathbf{m}-\mathbf{m}'} E_{\beta\mathbf{m}'}. \qquad (4)$$

Here vector index $\mathbf{m} = (m_1, m_2, m_3)$ enumerates Fourier harmonics, and possible $\mathbf{k}$ component values are $k_{\alpha\mathbf{m}} = k_\alpha^{inc} + m_\alpha K_\alpha$, $K_\alpha = 2\pi/\Lambda_\alpha$, $m_\alpha, \alpha = 1,2,3$. The source of external field $E_{\alpha\mathbf{m}}^{inc}$ can be, for example, either a beam refracted at composite structure boundary or dipoles inside the structure. Within the scope of linear theory, the field generated by such source can be decomposed into plane wave spectrum. Therefore, we can take a plane wave in form $E_{\alpha\mathbf{m}}^{inc} = \delta_{\mathbf{m}0}a_\alpha^{inc}\exp(i\mathbf{k}^{inc}\mathbf{r})$ with some wavevector $\mathbf{k}^{inc}$ as the incident field, without loss of generality. Eq. (4) has no singular terms since here and further $\varepsilon_b$ is supposed to be chosen so that $k_\mathbf{m}^2 \neq k_b^2$ for any $\mathbf{m}$ (which is always possible due to denumerability of $\mathbf{m}$ set). Note that the only wavevector governed by the dispersion equation here is $\mathbf{k_0} = \mathbf{k}^{inc}$, so that $\left|\mathbf{k}^{inc}\right|^2 = \omega^2 n_i^2 \mu_0$, where $n_i^2$ will be a variable in the mode propagation search procedure by means of resonance analysis of solutions to Eq. (4). An analogous equation system for unknown magnetic field is derived similarly and writes

$$H_{\alpha \mathbf{m}} = H_{\mathbf{m}\alpha}^{inc} + \sum_{\beta,\gamma=1,2,3} \xi_{\alpha\beta\gamma} \frac{k_{\beta \mathbf{m}}}{k_b^2 - k_{\mathbf{m}}^2} \sum_{\mathbf{m}'} \left[ \delta_{\mathbf{mm}'} - \left(\frac{\varepsilon_b}{\varepsilon}\right)_{\mathbf{m}-\mathbf{m}'} \right] \sum_{\sigma,\upsilon=1,2,3} \xi_{\gamma\sigma\upsilon} k_{\sigma \mathbf{m}'} H_{\upsilon \mathbf{m}'} , \qquad (5)$$

where $\xi_{\alpha\beta\gamma}$ is Levi-Civita symbol that defines vector product components in Cartesian coordinates.

A finite linear system is obtained from either Eq. (4) or Eq. (5) by truncation. Fourier series truncation is straightforward when coordinate functions $\mathbf{E}(\mathbf{r})$ and $\varepsilon(\mathbf{r})$ are smooth. In case of discontinuous permittivity and fields, these equations should be reformulated to avoid products of discontinuous functions with coinciding points of discontinuity before the truncation procedure. This is analogous to what is usually done in the Fourier methods in the diffraction grating theory [17,22]. This reformulation can be performed in 3D periodic structure case also. However, in the diffraction grating theory, the use of such mathematically incorrect equation affects the convergence of the method rather than the verity of obtained solutions, and is not critical for pure dielectric structures. Therefore, in this article we rest upon truncated Eqs. (4) and (5), since we are aimed at presenting the method rationale. We aim at presenting fine tuning features in a later publication. As it will be seen further even these simplified formulations exhibit quite acceptable convergence for effective parameters retrieval in dielectric structures.

The truncation procedure takes into account only $N_x$, $N_y$, and $N_z$ Fourier harmonics along axes *X*, *Y*, and *Z* respectively, and, e.g., transforms Eq. (4) into a finite matrix-vector equation on unknown electric field component amplitude Fourier images. Its solution by an iterative Krylov subspace method like the Stabilized Bi-Conjugatre Gradient (BiCGSTAB) or the Generalized Minimal Residual (GMRes) is particularly efficient since the matrix to be inverted can be multiplied by a vector in a fast manner (we make use of the GSMRes, which is rationalized below). The heaviest procedure is the multiplication of matrix $(\Delta\varepsilon/\varepsilon_b)_{\mathbf{m}-\mathbf{m}'}$ by a vector which can be accelerated by the Fast Fourier Transform since Fourier matrix $(\Delta\varepsilon/\varepsilon_b)_{\mathbf{m}-\mathbf{m}'}$ has Toeplitz structure. Note that we do not make a transform to the coordinate space where multiplication by permittivity is diagonal. Instead, we prefer to benefit from analytically found permittivity Fourier components (Fourier images of sphere and polyhedron can be combined to represent complex structures). Multiplications define $O(N\log N)$ overall algorithm computer time and $O(N)$ memory consumption by the whole solution procedure, where $N = N_x N_y N_z$. Details of the fast multiplication can be found, e.g., in previous publications on the GSM [18,19].

Modes in 3D crystals

Equation (4) (or analogous equation for the magnetic field) being solved provides an amplitude vector for each Fourier harmonic of the field. When an incident wavevector modulo being proportional to $n_i$ at a given frequency approaches the mode propagation constant in the given direction the field response exhibits resonant behavior. This resonance is described algebraically by an isolated complex pole. Then, in order to calculate propagation constants by the use of rigorous solutions obtained in accordance with the previous section we represent the 0-th Fourier order response field as a meromophic function of incident field wavevector modulo $\kappa = \left|\mathbf{k}^{inc}\right|$:

$$\xi(\kappa) = \sum_{j=1}^{N_p} \frac{\zeta_j}{\kappa - \eta_j} + \sum_{j=0}^{\infty} \gamma_j \kappa^j \tag{6}$$

The term $\xi(\kappa)$ stands for either electric or magnetic field $0^{th}$ order vector component amplitude, propagation constants $\eta_j$ are simple poles of function $\xi(\kappa)$ with modal field amplitudes $\zeta_j$, and constant vector coefficients $\gamma_j$ represent the regular part of the filed decomposition. In fact, any diffraction order possesses a similar resonant behavior. The choice of the $0^{th}$ order here is the most appropriate since in the effective medium simulation this order has usually the largest amplitude which would allow for decreasing numerical accuracy loss.

If one truncates infinite series in (6) the equation becomes a rational approximation of the field response. Formally, it resembles the Pade approximation [23] which is widely used in spectral approximations [24,25]. However, contrary to the Pade approximation which is designed to reproduce analytic functions, we are interested in a decomposition which better approximates a meromorphic function near its poles. Thus, in this work we focus primarily on the most accurate search of complex poles $\eta_j$ and amplitudes $\zeta_j$, and therefore, apply the algorithm based on the numerical derivation [26] instead of commonly used Baker's algorithm designed for the best function approximation in a given spectral interval.

Briefly, the algorithm based on the numerical derivation [26] is as follows. First, a search of propagation constants $\eta_j$ is reduced to determine polynomial coefficients $p_k$:

$$P_{N_p}(\kappa) \equiv \prod_{j=1}^{N_p} (\kappa - \eta_j) = \sum_{k=0}^{N_p} p_k \kappa^k \tag{7}$$

Then, product $P_{N_p}(\kappa)\xi(\kappa)$ is an analytic function, and applying to it the $N_d$-th order Newton divided difference operator in some interval $\kappa_{min} \leq \kappa \leq \kappa_{max}$ containing $N_d + 1$ points $\kappa_p$, $(N_d > N_p)$ one gets an approximate equality

$$D_{N_d}\left[P_{N_p}(\kappa)\xi(\kappa)\right]\!\kappa_0,\ldots,\kappa_{N_d}\approx 0. \tag{8}$$

Using the explicit expression for the divided difference and choosing $N_p - 1$ independent point sets at interval $\kappa_{min} \leq \kappa \leq \kappa_{max}$ one obtains linear equation system for unknown coefficients $p_k$:

$$\sum_{k=0}^{N_p-1}\left[\sum_{i=0}^{N_d}\frac{\kappa_i^k}{\xi_i\prod_{\substack{j=0\\j\neq i}}^{N_d}(\kappa_i-\kappa_j)}\right]p_k = -\sum_{i=0}^{N_d}\frac{\kappa_i^{N_p}}{\xi_i\prod_{\substack{j=0\\j\neq i}}^{N_d}(\kappa_i-\kappa_j)}. \tag{9}$$

Zeros $\eta_j$ are found using a standard QR factorization procedure as eigenvalues of the companion matrix of polynomial (7). With propagation constants $\eta_j$ being found explicitly, calculation of pole amplitudes becomes straightforward:

$$\zeta_k = \left[\prod_{\substack{i=1\\i\neq k}}^{N_p}(\eta_k-\eta_i)\right]^{-1}\sum_{i=0}^{N_d}\xi_i\prod_{j=1}^{m}(\kappa_i-\eta_j)\prod_{j\neq i}^{N_d}\frac{(\eta_k-\kappa_j)}{(\kappa_i-\kappa_j)}, k=1,\ldots,N_p. \tag{10}$$

The described algorithm has two parameters: number of poles $N_p$, and number of divided difference operator points $N_d$. Although the propagation in each direction in an effective medium can be described by two refractive indices (solutions to the Fresnel normal equation), it is preferable to choose $N_p$ to be greater than 2 for numerical reasons. A good choice of $N_p$ and $N_d$ values for the dielectric tensor calculation was found to be $N_p \sim 5$, $N_d \sim 10$. In this case, pole calculation accuracy is high enough to not affect the convergence of effective dielectric tensor components.

## Calculation details

An essential part of the method is evaluation of the dielectric permittivity Fourier images. We prefer to benefit from analytically known Fourier images of simple shapes as sphere or polyhedron. These exact images can be combined to obtain Fourier matrices of most experimentally realizable protonic crystal structures. Note also that Fourier components of inverse permittivity $[\varepsilon_b/\varepsilon]_m$ within the H-field formulation are found analytically just in the same way as matrix $[\varepsilon/\varepsilon_b]$ components, so there is no need to invert numerically matrix $[\varepsilon/\varepsilon_b]$. Such technique differs from most Fourier-based approaches implementing a diagonal coordinate space permittivity by field multiplication, where permittivity is evaluated at some spatial mesh points. Implementation of a finite mesh introduces inaccuracy which in turn requires an additional effort to overcome it by some artificial boundary averaging procedure (e.g., [14]).

From computational viewpoint, the described method relies mainly on two basic numerical procedures: the GMRes and the Fast Fourier Transform. These two define primarily program time and memory consumption. The reason for using the GMRes instead of, e.g., the BiCGSTAB consists in fact, that the GMRes converges at each point of an equidistant mesh $\kappa_p$, $0 \leq p \leq N_d$ even for points being close to poles, whereas other iterative methods fail to converge in the vicinity of poles. Nevertheless, other matrix inversion iterative methods probably could be implemented in the described algorithm, however, this issue lies beyond the scope of the paper and will be discussed elsewhere. Number of GMRes iterations in all examples presented further varied in range from about 10 for low-contrast structures to about 50 for high-contrast structures.

Program based on the described method was implemented for both CPU and GPU calculations. Both programs were written under Windows 7 platform using Microsoft Visual Studio 2010 Professional edition. The GPU program was written in CUDA language. The GMRes algorithm was implemented according to Saad [27] with the modified Gram-Schmidt orthogonalization. Internal vector operations were implemented via CUBLAS procedures. For the FFT we used the Cooley-Tukey algorithm. For GPU computation we used our own FFT implementation which algorithmically is very close to the well-known Volkov FFT code [28] but was optimized for the double precision arithmetic.

Calculations were performed on Intel Xeon E5640 processor, and on graphic card GeForce GTX Titan with about 6 Gb of memory. Calculation time comparison example for CPU and GPU simulations is shown in Table 1 (each time is an average over 10 procedure runs). The table demonstrates a tremendous gain provided by the GPU when the Fourier harmonic number exceeds several thousands. It is this gain that becomes a cornerstone of the described rigorous homogenization method and enables extremely efficient effective dielectric tensor rigorous computation. Even when the GPU memory is insufficient to store all GMRes vectors, and an intense memory exchange between GPU memory and RAM, the GPU program version appears to be more than 30 times faster than its CPU counterpart.

Table 1. Comparison of calculation time for the GSM rigorous 3D diffraction simulation on CPU and on GPU (for 3D lattice of spheres having refractive index 3 in the air). Each time value is averaged over 10 corresponding procedure runs. For 256×256×256 Fourier orders 6 GB of GPU memory is not enough to store sufficient number of GMRes vectors, and an intense memory exchange between GPU and RAM is required. GMRes converged to $10^{-8}$ accuracy at about 30 iterations.

| Number of Fourier orders | 16×16×16 | 32×32×32 | 64×64×64 | 128×128×128 | 256×256×256 |
|---|---|---|---|---|---|
| CPU-version time, s | 0.515 | 4.204 | 39.047 | 409.709 | 3938.634 (~1 hour) |
| GPU-version time, s | 0.166 | 0.244 | 0.448 | 2.184 | 113.676 (~2 minutes) |
| Time gain | ~3 | ~17 | ~87 | ~188 | ~35 |

Numerical convergence

By means of the developed codes, we investigated numerical convergence of the method for different types of lattices. Namely, mode propagation constants and modal field amplitudes were calculated for different number of Fourier harmonics $N_O$, and an absolute difference of these modal parameters versus the number of the Fourier harmonic was traced. Figures 2-3 demonstrate the obtained convergence for two types of 3D crystals (cubic and spherical air holes in the host material) with different refractive index contrast for both E- and H-formulations. Refractive index and field amplitude differences demonstrated a very close behavior, their representative values are laid off along the vertical axis. The figures show that an accuracy of order of $10^{-5}$ in the calculated refractive index can be achieved even for high contrast structures.

Comparing the E- and the H-formulations it is seen that both possess a similar convergence rate, whereas the former provides generally several times better accuracy. This demonstrates an important advantage before the conventional rigorous methods [11], where the electric field formulation leads to substantial decrease of convergence.

Superiority of the proposed method over quasistatic approaches can be demonstrated by studying effective parameters of 3D periodic composite structures while increasing structure periods. As an example here we consider photonic crystals with effective uniaxial and biaxial behavior. The photonic crystals under consideration are spherical air cavities inside a homogeneous matrix with refractive index 1.5 periodically located along the Cartesian coordinate axes (as in Fig. 1). In the limit of zero wavelength-to-period ratio, uniaxial structure behavior is achieved by letting a period along one axis to be different than the other equal two, whereas the biaxial behavior is observed when all three periods are different. These lattice types are conventionally referred to as simple tetragonal and simple orthorhombic, respectively. Note that due to the structure symmetry the principal refractive index axes coincide in these cases with the Cartesian coordinate axes.

In order to compare simulation results for different periods with a quasistatic solution, which is inherently independent of period, we fix the structure volume filling factor and alternate structure periods. Under this assumption, the cavity radius would change proportionally to periods. Consider first simple tetragonal lattice crystal with uniaxial effective behavior. Set the sphere radius to be $1/3$ of the period when all three periods are equal, so that the volume filling factor is about $f = 0.155$, and vary the period along X axis $\Lambda_x$ under conditions $f = const$, and $\Lambda_y = \Lambda_z$. Fig. 4a shows the dependence of ordinary and extraordinary effective refractive indices on the period-to-wavelength ratio $\Lambda_x/\lambda$ for increasing cavity radius (0.01λ, 0.02λ, 0.03λ, 0.04λ, and 0.05λ). Black solid line shows the Maxwell-Garnett quasistatic solution equal to $n_{MG,uniaxial} = 1.4213$. The well-known effect of the effective refractive index increase for finite period to wavelength ratio in comparison with the Maxwell-Garnett result can be seen along vertical line $\Lambda_x/\Lambda_{y,z} = 1$, which corresponds to an effective isotropic crystal. For other values of $\Lambda_x/\lambda$ the anisotropy causes the propagation constant split, and one of the effective indices for certain directions can become lower than that in the quasistatic limit. Simultaneously, the period increase equally shifts both curves corresponding to ordinary and extraordinary indices in the direction of higher propagation constants. Fig. 4b demonstrates corresponding change of the Fresnel normal surface in form of shifted cross sections of this surface by coordinate planes. The blue solid line shows the cross section for $\Lambda_x/\Lambda_{y,z} = 0.7$ which is in the left hand part of Fig. 4a, whereas cross-section shown by the red dashed line is drawn for $\Lambda_x/\Lambda_{y,z} = 1.3$ – the right-hand part of Fig. 4a.

Graphs in Fig. 5 are similar to those of Fig. 4 but correspond to orthorhombic lattice type crystal with biaxial effective behavior. Here we fixed ratio $\Lambda_y/\Lambda_z = 1.1$ and alternated ratio

$\Lambda_x/\Lambda_y$. The sphere radius was chosen $\Lambda_x/3$ under condition $\Lambda_x = \Lambda_z$, so the filling factor was $f = 0.141 = const$. The effective refractive indices $n_{x,y,z}$ along dielectric tensor principal axes versus $\Lambda_x/\Lambda_z$ ratio are plotted in Fig. 5a for several different cavity radii (0.01λ, 0.02λ, 0.03λ, 0.04λ, and 0.05λ). It is seen that bianisotropy allows to achieve either superiority of all three principal refractive indices over the Maxwell-Garnett solution ($n_{MG,biaxial} = 1.4284$) or to make two of them lower than $n_{MG,biaxial}$. The Fresnel surface cross-sections corresponding to the left-hand and the right-hand parts of the graph in Fig. 5a are shown in Fig. 5b demonstrating a shift of the effective crystal optical axis from coordinate plane XY to plane XZ.

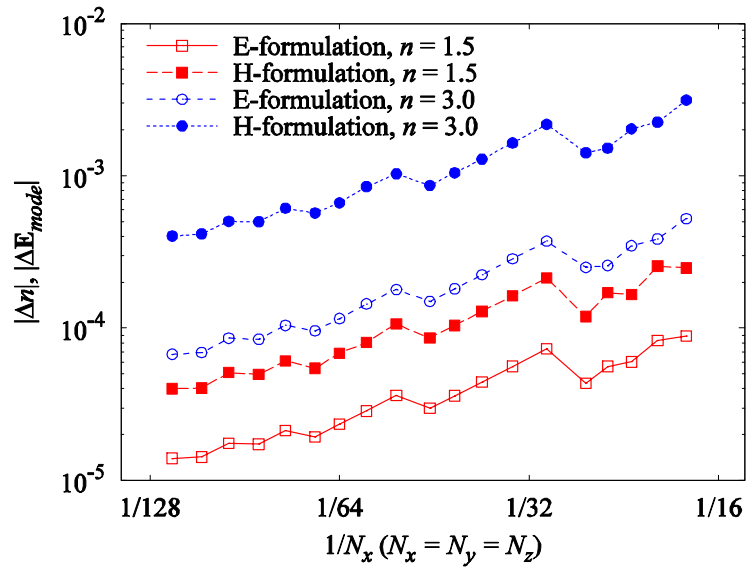

Figure 2. Convergence of the mode propagation constant and the modal field for the E-formulation and the H-formulation in a crystal being an infinite 3D periodic set of air cubic cavities in a medium with refractive index 1.5 and 3.0. The periods along all three coordinate axes are the same and equal to 0.3λ, and the cube side length is half of the period. The calculated effective index is 1.441688 for the host medium with $n = 1.5$ and 2.82503 for the host medium with $n = 3.0$.

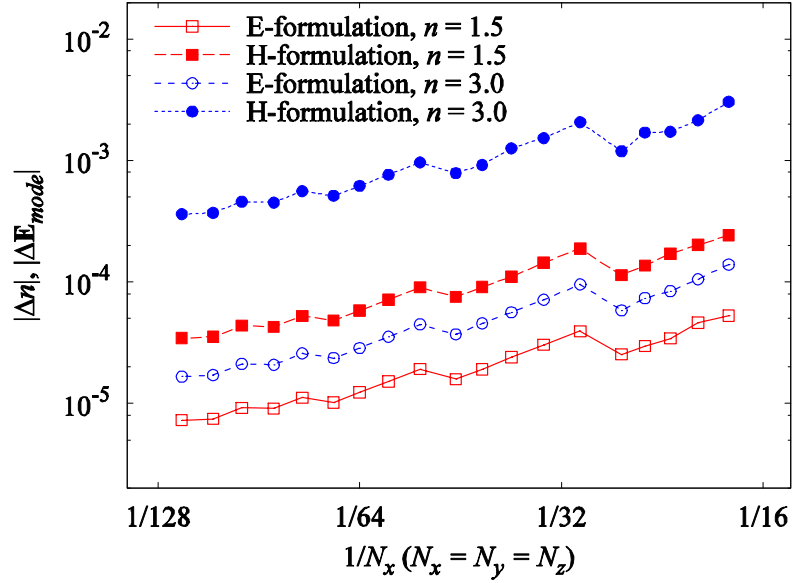

Figure 3. Convergence of the mode propagation constant and the modal field for the E-formulation and the H-formulation in a crystal being an infinite 3D periodic set of air spherical cavities in media with refractive indices 1.5 and 3.0. The periods along all three coordinate axes are the same and equal to 0.3λ, the sphere radius is 0.125λ. The calculated effective index equals 1.359786 for the host medium with $n = 1.5$ and 2.53781 for the host medium with $n = 3.0$.

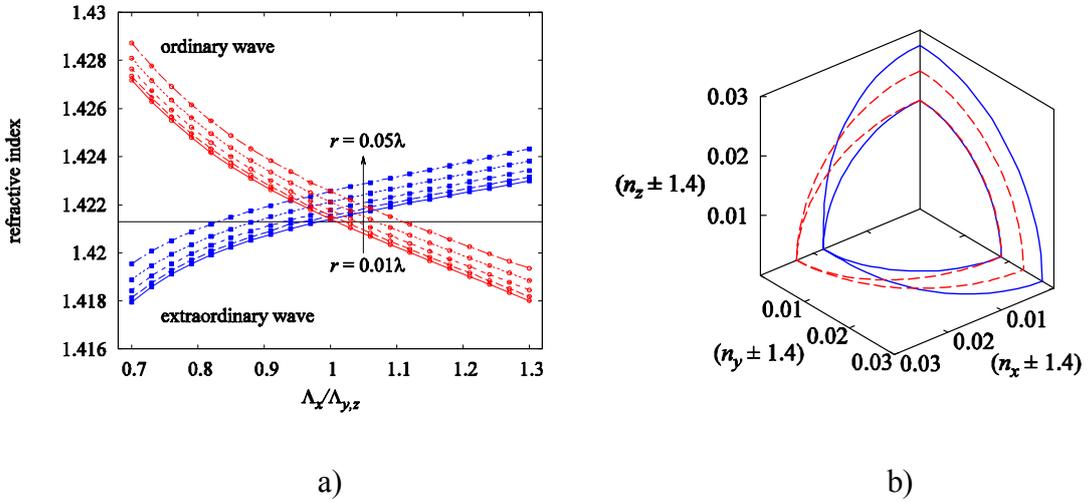

a)  b)

Fig. 4. a) Ordinary and extraordinary effective indices of a crystal represented by a tetragonal lattice of spherical cavities in a homogeneous matrix with $n = 1.5$ versus the period ratio for increasing cavity radius (0.01λ, 0.02λ, 0.03λ, 0.04λ, and 0.05λ) and the constant volume filling factor. The sphere radius is 1/3 of the period whereas all three periods are equal. Maxwell-Garnett solution corresponding to the given filling factor is shown by the black solid horizontal line. b) Cross-sections of the shifted normal surface by coordinate planes for the described effective uniaxial crystal corresponding to the cavity radius 0.02λ. The blue solid line corresponds to period ratio $\Lambda_x/\Lambda_{y,z} = 0.7$, the red dashed line corresponds to $\Lambda_x/\Lambda_{y,z} = 1.3$.

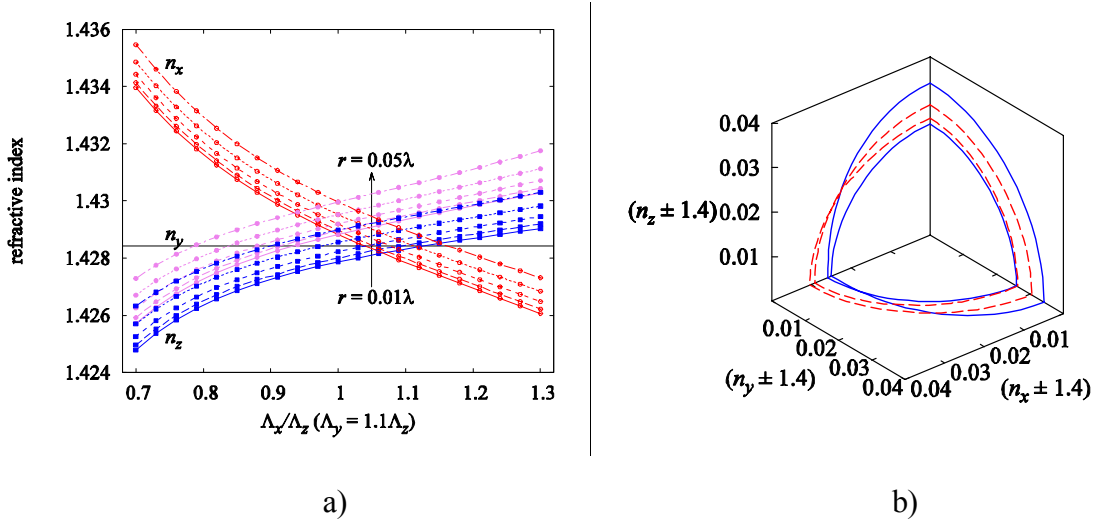

a) b)

Fig. 5. a) Effective indices along principal axes of a crystal represented by an orthorhombic lattice of spherical cavities in a homogeneous matrix with $n = 1.5$ versus the period ratio $\Lambda_x/\Lambda_z$ under fixed ratio $\Lambda_y/\Lambda_z = 1.1$ for increasing cavity radius (0.01λ, 0.02λ, 0.03λ, 0.04λ, and 0.05λ) and the constant volume filling factor. The red curve with circle points corresponds to the diagonal refractive index tensor x-component, the magenta curve with filled circles – to the y-component, and the blue curve with square points – to the z-component. The sphere radius is $\Lambda_x/3$ of the period with $\Lambda_x = \Lambda_z$. Maxwell-Garnett solution corresponding to the given filling factor is shown by the black solid horizontal line. b) Cross-sections of the shifted normal surface by coordinate planes for the described effective uniaxial crystal corresponding to cavity radius 0.02λ. The blue solid line corresponds to period ratio $\Lambda_x/\Lambda_z = 0.7$, the red dashed line – to $\Lambda_x/\Lambda_z = 1.3$.

Conclusion

The article presents the new Fourier space based approach to retrieving effective permittivity tensor in 3D periodic dielectric composites. The new method relies upon the GSM developed previously for the 1D and 2D diffraction grating simulation and retains the linear complexity relative to the number of Fourier orders. It includes two principal steps. The first step consists in rigorous electromagnetic solution to the 3D diffraction problem. Obtained rigorous solutions are then used at the second step fin the analysis of structure resonant response yielding mode propagation constants and modal fields. The mode parameter search was shown to provide resulting accuracy about $10^{-5}$ and even better for a wide range of refractive index variations. At the same time, GPU enabled computations allow for extremely fast analysis of complex crystals which may open a promising way to inverse problem solution concerning effective behavior of complex photonic structures. A future work should include reformulation of the method within the scope of mathematically correct treatment of distribution products and a more detailed investigation of its potential.

Acknowledgements

This work was supported by the Ministry of Education and Science of the Russian Federation (no. 16.19.2014/K), Russian Foundation for Basic Reseatch (no. NK 14-07-31352\14), and Russian President Grant for Young Scientists (no. MK-6303.2015.9).